\documentclass[fleqn,11pt,twoside]{article}

\usepackage{amsthm,amsthm,amssymb, color, xcolor,epsfig, graphics, subfigure}

\usepackage{amsmath, graphicx, latexsym, lscape }

\makeatletter
\newcommand{\copyrightnote}[2]{{\renewcommand{\thefootnote}{}
 \footnotetext{\small\it
\begin{flushleft}
 \copyright \ #1   #2
\end{flushleft}}}}

\newcommand{\Name}[1]{\begin{flushleft}
                       \LARGE \bf #1
                       \end{flushleft}\vspace{-3mm}}

\newcommand{\Author}[1]{\begin{flushleft}
                       \it #1 \end{flushleft}}

\newcommand{\Address}[1]{\begin{flushleft}
                       \it #1 \end{flushleft}}

\newcommand{\Date}[1]{\begin{flushleft}
                      \small  \it #1 \end{flushleft}}

\newcommand{\evenhead}{Author \ name}
\newcommand{\oddhead}{Article \ name}

\renewcommand{\@evenhead}{
\hspace*{-3pt}\raisebox{-15pt}[\headheight][0pt]{\vbox{\hbox to \textwidth
{\thepage \hfil \evenhead}\vskip4pt \hrule}}}
\renewcommand{\@oddhead}{
\hspace*{-3pt}\raisebox{-15pt}[\headheight][0pt]{\vbox{\hbox to \textwidth
{\oddhead \hfil \thepage}\vskip4pt\hrule}}}
\renewcommand{\@evenfoot}{}
\renewcommand{\@oddfoot}{}

\long\def\@makecaption#1#2{%
  \vskip\abovecaptionskip
  \sbox\@tempboxa{\small \textbf{#1.}\ \ #2}%
  \ifdim \wd\@tempboxa >\hsize
    {\small \textbf{#1.}\ \ #2}\par
  \else
    \global \@minipagefalse
    \hb@xt@\hsize{\hfil\box\@tempboxa\hfil}%
  \fi
  \vskip\belowcaptionskip}

\newcommand{\JNMPnumberwithin}[3][\arabic]{%
  \@ifundefined{c@#2}{\@nocounterr{#2}}{%
    \@ifundefined{c@#3}{\@nocnterr{#3}}{%
      \@addtoreset{#2}{#3}%
      \@xp\xdef\csname the#2\endcsname{%
        \@xp\@nx\csname the#3\endcsname .\@nx#1{#2}}}}%
}

\newcommand{\resetfootnoterule} {
  \renewcommand\footnoterule{%
  \kern-3\p@
  \hrule\@width.4\columnwidth
  \kern2.6\p@}
}

\renewcommand{\footnoterule}{}

\makeatother

\theoremstyle{definition}

\begin{document}

\renewcommand{\evenhead}{ {\LARGE\textcolor{blue!10!black!40!green}{{\sf \ \ \ ]ocnmp[}}}\strut\hfill Maciej B\l aszak}
\renewcommand{\oddhead}{ {\LARGE\textcolor{blue!10!black!40!green}{{\sf ]ocnmp[}}}\ \ \ \ \   Nonhomogeneous DWW and Painlev\'{e}
equations}

\thispagestyle{empty}
\newcommand{\FistPageHead}[3]{
\begin{flushleft}
\raisebox{8mm}[0pt][0pt]
{\footnotesize \sf
\parbox{150mm}{{Open Communications in Nonlinear Mathematical Physics}\ \ \ {\LARGE\textcolor{blue!10!black!40!green}{]ocnmp[}}
\ \ Vol.1 (2021) pp
#2\hfill {\sc #3}}}\vspace{-13mm}
\end{flushleft}}

\FistPageHead{1}{\pageref{firstpage}--\pageref{lastpage}}{ \ \ Letter}

\strut\hfill

\strut\hfill

\copyrightnote{The author(s). Distributed under a Creative Commons Attribution 4.0 International License}

\qquad\qquad\qquad\qquad\qquad\qquad {\LARGE  {\sf Letter to the Editors}}

\strut\hfill

\Name{Nonhomogeneous Dispersive Water Waves and Painlev\'{e}
equations}

\Author{Maciej B\l aszak}

\Address{Faculty of Physics, Department of Mathematical Physics
and Computer Modelling,\\ A. Mickiewicz University, Uniwersytetu
Pozna\'nskiego 2, 61-614 Pozna\'{n}, Poland}

\Date{Received Date June 18, 2021; Accepted Date July 21, 2021}

\setcounter{equation}{0}

\begin{abstract}
\noindent In this letter we consider three nonhomogeneous
deformations of Dispersive Water Wave (DWW) soliton equation and
prove that their stationary flows are equivalent to three famous
Painlev\'{e} equations, i.e. $P_{II}$, $P_{III}$ and $P_{IV},$
respectively.
\end{abstract}

\label{firstpage}

The six classical Painlev\'{e} equations ($P_{I}-P_{VI}$) are
integrable nonlinear second-order ordinary differential equations
(ODE's) that defined new transcendental functions globally in the
complex plane. Although first
discovered from strictly mathematical considerations, nowadays the Painlev%
\'{e} equations play an important role in a variety of physical
applications. In particular includes such topics as: the transport
of particles across boundaries (Nernst--Planck equations)
\cite{A1}, Hele-Shaw problems in viscous fluids \cite{A2}, dilute
Bose--Einstein condensates in an external one dimensional field
(Gross--Pitaevskii equation) \cite{A4} as well as various
approaches to quantum field theory, statistical mechanics, plasma
physics, nonlinear waves, quantum gravity, and nonlinear fiber
optics. Besides, in last decades there has been considerable
interest in the Painlev\'{e} equations also due to the fact that
they arise as various reductions of the nonlinear soliton PDE's,
which are solvable by inverse scattering method. In the literature
they mainly appeared as similarity reductions and scaling
reductions of particular soliton equations (see for example
\cite{C}-\cite{Ar} and references therein).

In this letter we announce the existence of another relationship
between soliton systems and Painlev\'{e} equations. Actually,
Painlev\'{e} equations are just equivalent to stationary flows of
particular, nonhomogeneous deformations of soliton systems. Here
we prove the equivalence between three Painlev\'{e} equations,
i.e.
\begin{equation}
P_{II}:\ \ \ \ \ q_{\tau \tau }=2q^{3}+\tau q+\alpha ,
\label{1.1}
\end{equation}%
\begin{equation}
P_{III}:\ \ \ \ \ \tau qq_{\tau \tau }=\tau q_{\tau }^{2}-qq_{\tau
}+\gamma \tau q^{4}+\alpha q^{3}+\beta q+\delta \tau  \label{1.2}
\end{equation}%
and
\begin{equation}
P_{IV}:\ \ \ \ \ qq_{\tau \tau }=\frac{1}{2}q_{\tau }^{2}+\frac{3}{2}%
q^{4}+4\tau q^{3}+2(\tau ^{2}-\alpha )q^{2}+\beta ,  \label{1.3}
\end{equation}%
where $\alpha ,\beta ,\gamma ,\delta $ are constants, and
stationary flows of the Dispersive Water Wave (DWW) PDE, deformed
by its own local master symmetries. Nevertheless, our preliminary
research shows that such relation
exists on the level of whole soliton hierarchies and multi component Painlev%
\'{e}-type equations.

Let us consider the DWW soliton equation in Antonowicz-Fordy
representation \cite{A}
\begin{equation}
\left(
\begin{array}{c}
u \\
v%
\end{array}%
\right) _{t}=\left(
\begin{array}{c}
\frac{1}{4}v_{xxx}+uv_{x}+\frac{1}{2}vu_{x} \\
u_{x}+\frac{3}{2}vv_{x}%
\end{array}%
\right) \equiv \mathcal{K}_{2}.  \label{2.0}
\end{equation}%
It is nonlinear PDE related to a linear spectral problem
\begin{equation*}
(\partial _{x}^{2}+u+v\lambda )\psi =\lambda ^{2}\psi .
\end{equation*}%
Equation (\ref{2.0}) belongs to tri-Hamiltonian soliton hierarchy
\cite{A}
\begin{equation}
\left(
\begin{array}{c}
u \\
v%
\end{array}%
\right) _{t_{n}}=\mathcal{K}_{n}=\pi _{0}\gamma _{n}=\pi
_{1}\gamma _{n-1}=\pi _{2}\gamma _{n-2},\ \ \ \ \ \ n=1,2,...
\label{2.2}
\end{equation}%
where $\gamma _{r}$ are exact one-forms and three Poisson
operators are
\begin{equation*}
\pi _{0}=\left(
\begin{array}{cc}
-\frac{1}{2}v\partial _{x}-\frac{1}{2}\partial _{x}v & \partial _{x} \\
\partial _{x} & 0%
\end{array}%
\right) ,\ \ \ \pi _{1}=\left(
\begin{array}{cc}
\frac{1}{4}\partial _{x}^{3}+\frac{1}{2}u\partial
_{x}+\frac{1}{2}\partial
_{x}u & 0 \\
0 & \partial _{x}%
\end{array}%
\right) ,\
\end{equation*}%
\begin{equation*}
\ \ \pi _{2}=\left(
\begin{array}{cc}
0 & \frac{1}{4}\partial _{x}^{3}+\frac{1}{2}u\partial _{x}+\frac{1}{2}%
\partial _{x}u \\
\frac{1}{4}\partial _{x}^{3}+\frac{1}{2}u\partial
_{x}+\frac{1}{2}\partial
_{x}u & \frac{1}{2}v\partial _{x}+\frac{1}{2}\partial _{x}v%
\end{array}%
\right) .
\end{equation*}%
The hierarchy (\ref{2.2}) can be generated by recursion operator
and its adjoint
\begin{equation*}
N=\pi _{1}\pi _{0}^{-1}=\left(
\begin{array}{cc}
0 & \frac{1}{4}\partial _{x}^{2}+u+\frac{1}{2}u_{x}\partial _{x}^{-1} \\
1 & v+\frac{1}{2}v_{x}\partial _{x}^{-1}%
\end{array}%
\right) ,\ \ \ N^{\dagger }=\left(
\begin{array}{cc}
0 & 1 \\
\frac{1}{4}\partial _{x}^{2}+u-\frac{1}{2}\partial _{x}^{-1}u_{x} & v-\frac{1%
}{2}\partial _{x}^{-1}v_{x}%
\end{array}%
\right) ,
\end{equation*}%
\begin{equation*}
\mathcal{K}_{n+1}=N^{n}\mathcal{K}_{1},\ \ \ \gamma _{n}=d\mathcal{H}%
_{n}=(N^{\dagger })^{n}\gamma _{0},\ \ \ \ n=0,1,2,...
\end{equation*}%
where $\mathcal{H}_{r}$ are Hamiltonian densities.\ In particular
\begin{equation*}
\gamma _{0}=\left(
\begin{array}{c}
2 \\
v%
\end{array}%
\right) ,~\ \ \gamma _{1}=\left(
\begin{array}{c}
v \\
u+\frac{3}{4}v^{2}%
\end{array}%
\right) ,\ \ \ \gamma _{2}=\left(
\begin{array}{c}
u+\frac{3}{4}v^{2} \\
\frac{1}{4}v_{xx}+\frac{3}{2}uv+\frac{5}{8}v^{3}%
\end{array}%
\right) ,\ \ ...
\end{equation*}%
\begin{equation*}
\ \mathcal{H}_{0}=2u+\frac{1}{2}v^{2},\ \ \ \ \mathcal{H}_{1}=uv+\frac{1}{4}%
v^{3},\ \ \ \mathcal{H}_{2}=-\frac{1}{8}v_{x}^{2}+\frac{1}{2}u^{2}+\frac{3}{4%
}uv^{2}+\frac{5}{32}v^{4},\ \ \ ...
\end{equation*}%
\begin{equation*}
\mathcal{K}_{1}=\left(
\begin{array}{c}
u_{x} \\
v_{x}%
\end{array}%
\right) ,\ \ \ \mathcal{K}_{2}=\left(
\begin{array}{c}
\frac{1}{4}v_{xxx}+uv_{x}+\frac{1}{2}vu_{x} \\
u_{x}+\frac{3}{2}vv_{x}%
\end{array}%
\right) ,\ \ \ ...
\end{equation*}%
In addition, with the DWW hierarchy of symmetries
$\mathcal{K}_{n}$ is related a hierarchy of master symmetries
$\sigma _{m}=N^{m+1}\sigma _{-1},$ non-local in general, except
the first three
\begin{equation*}
\sigma _{-1}=\left(
\begin{array}{c}
-v \\
2%
\end{array}%
\right) ,\ \ \ \sigma _{0}=\left(
\begin{array}{c}
2u+xu_{x} \\
v+xv_{x}%
\end{array}%
\right) ,\ \ \ \sigma _{1}=\left(
\begin{array}{c}
\frac{3}{4}v_{xx}+\frac{1}{4}xv_{xxx}+uv+xuv_{x}+\frac{1}{2}xvu_{x} \\
xu_{x}+\frac{3}{2}xvv_{x}+v^{2}+2u%
\end{array}%
\right) ,\ \ \ ...
\end{equation*}%
Notice that
\begin{equation*}
\sigma _{-1}=\pi _{0}\zeta ,\ \ \ \sigma _{0}=\pi _{1}\zeta ,\ \ \
\sigma _{1}=\pi _{2}\zeta ,\ \ \ \zeta =\left(
\begin{array}{c}
2x \\
xv%
\end{array}%
\right) .
\end{equation*}%
Both, symmetries $\mathcal{K}_{n}$ and master symmetries $\sigma
_{m}$ constitute so called Virasoro algebra (hereditary algebra)
\begin{equation*}
\lbrack \mathcal{K}_{m},\mathcal{K}_{n}]=0,\ \ \ [\mathcal{\sigma }_{m},%
\mathcal{K}_{n}]=n\mathcal{K}_{n+m},\ \ \ \ [\sigma _{m},\sigma
_{n}]=(n-m)\sigma _{n+m}.
\end{equation*}

Let us consider three Hamiltonian nonhomogeneous deformations of
the DWW
equation (\ref{2.0}) by its local master symmetries%
\begin{equation}
\left(
\begin{array}{c}
u \\
v%
\end{array}%
\right) _{t}=\mathcal{K}_{2}+\sigma _{-1}=\pi _{0}(\gamma
_{2}+\zeta )=\left(
\begin{array}{c}
\frac{1}{4}v_{xxx}+uv_{x}+\frac{1}{2}vu_{x}-v \\
u_{x}+\frac{3}{2}vv_{x}+2%
\end{array}%
\right) ,  \label{2.12a}
\end{equation}%
\begin{equation}
\left(
\begin{array}{c}
u \\
v%
\end{array}%
\right) _{t}=\mathcal{K}_{2}+\sigma _{0}=\pi _{1}(\gamma
_{1}+\zeta )=\left(
\begin{array}{c}
\frac{1}{4}v_{xxx}+uv_{x}+\frac{1}{2}vu_{x}+2u+xu_{x} \\
u_{x}+\frac{3}{2}vv_{x}+v+xv_{x}%
\end{array}%
\right) ,  \label{2.12b}
\end{equation}%
\begin{align}
\left(
\begin{array}{c}
u \\
v%
\end{array}%
\right) _{t}=& \mathcal{K}_{2}+\sigma _{1}=\pi _{2}(\gamma
_{0}+\zeta )
\label{2.12c} \\
=& \left(
\begin{array}{c}
\frac{1}{4}v_{xxx}+uv_{x}+\frac{1}{2}vu_{x}+\frac{3}{4}v_{xx}+\frac{1}{4}%
xv_{xxx}+uv+xuv_{x}+\frac{1}{2}xvu_{x} \\
u_{x}+\frac{3}{2}vv_{x}+xu_{x}+\frac{3}{2}xvv_{x}+v^{2}+2u%
\end{array}%
\right) .  \notag
\end{align}%
In what follows, we will show the equivalence between stationary
flows, i.e. $t=0$, of equations (\ref{2.12a})-(\ref{2.12c}) and
$P_{II},\ P_{IV},\ P_{III}$, respectively.

First, let us consider the stationary flow of equation (\ref{2.12a})%
\begin{equation*}
0=\left(
\begin{array}{cc}
-\frac{1}{2}v\partial _{x}-\frac{1}{2}\partial _{x}v & \partial _{x} \\
\partial _{x} & 0%
\end{array}%
\right) \left(
\begin{array}{c}
u+\frac{3}{4}v^{2}+2x \\
\frac{1}{4}v_{xx}+\frac{3}{2}uv+\frac{5}{8}v^{3}+xv%
\end{array}%
\right) .
\end{equation*}%
It is equivalent to the pair of equations
\begin{align*}
0& =-\left( \frac{1}{2}v\partial _{x}+\frac{1}{2}\partial
_{x}v\right)
\left( u+\frac{3}{4}v^{2}+2x\right) +\left( \frac{1}{4}v_{xx}+\frac{3}{2}uv+%
\frac{5}{8}v^{3}+xv\right) _{x}, \\
0& =\left( u+\frac{3}{4}v^{2}+2x\right) _{x}.
\end{align*}%
Integrating them once we find
\begin{align*}
0& =u+\frac{3}{4}v^{2}+2x+c, \\
0& =\frac{1}{2}c_{1}v_{x}+\frac{1}{4}v_{xx}+\frac{3}{2}uv+\frac{5}{8}%
v^{3}+xv-2\alpha .
\end{align*}%
Eliminating $u$ we get a single equation
\begin{equation*}
\frac{1}{4}v_{xx}=\frac{1}{2}v^{3}+cv+2xv+2\alpha ,
\end{equation*}%
which after rescaling
\begin{equation*}
v=2q,\ \ \ \ x=\frac{1}{2}\tau
\end{equation*}%
turns into
\begin{equation*}
q_{\tau \tau }=2q^{3}+(c+\tau )q+\alpha ,
\end{equation*}%
which is $P_{II}$ (\ref{1.1}) with the choice $c=0.$

Second, let us consider the stationary flow of equation (\ref{2.12b})%
\begin{equation*}
0=\left(
\begin{array}{cc}
\frac{1}{4}\partial _{x}^{3}+\frac{1}{2}u\partial
_{x}+\frac{1}{2}\partial
_{x}u & 0 \\
0 & \partial _{x}%
\end{array}%
\right) \left(
\begin{array}{c}
v+2x \\
u+\frac{3}{4}v^{2}+xv%
\end{array}%
\right) .
\end{equation*}%
Again, it is equivalent to the pair of equations
\begin{align*}
0=& \left( \frac{1}{4}\partial _{x}^{3}+\frac{1}{2}u\partial _{x}+\frac{1}{2}%
\partial _{x}u\right) \left( v+2x\right) , \\
0=& \left( u+\frac{3}{4}v^{2}+xv\right) _{x}.
\end{align*}%
Denote $\gamma =v+2x$ and multiply the first equation by $\gamma
.$ Then, we find
\begin{align*}
0& =\gamma \left( \frac{1}{4}\partial _{x}^{3}+\frac{1}{2}u\partial _{x}+%
\frac{1}{2}\partial _{x}u\right) \gamma =\left( \gamma \gamma _{xx}-\frac{1}{%
2}\gamma _{x}^{2}+u\gamma ^{2}\right) _{x}, \\
0& =\left( u+\frac{3}{4}v^{2}+xv\right) _{x}.
\end{align*}%
Integrating them once we get%
\begin{align*}
0& =\gamma \gamma _{xx}-\frac{1}{2}\gamma _{x}^{2}+u\gamma ^{2}-\beta , \\
0& =u+\frac{3}{4}v^{2}+xv-c.
\end{align*}%
Eliminating $u$ we derive a single equation
\begin{equation*}
\gamma \gamma _{xx}=\frac{1}{2}\gamma _{x}^{2}+\frac{3}{4}\gamma
^{4}-2x\gamma ^{3}+(x^{2}-c)\gamma ^{2}+\beta ,
\end{equation*}%
which after rescaling
\begin{equation*}
\gamma =-2^{\frac{1}{4}}q,\ \ \ \ x=2^{\frac{1}{4}}\tau ,\ \ 2^{-\frac{1}{2}%
}c=\alpha
\end{equation*}%
turns into $P_{IV}$ (\ref{1.3}).

Finally, let us consider the stationary flow of equation (\ref{2.12c})%
\begin{equation}
0=\left(
\begin{array}{cc}
0 & \frac{1}{4}\partial _{x}^{3}+\frac{1}{2}u\partial _{x}+\frac{1}{2}%
\partial _{x}u \\
\frac{1}{4}\partial _{x}^{3}+\frac{1}{2}u\partial
_{x}+\frac{1}{2}\partial
_{x}u & \frac{1}{2}v\partial _{x}+\frac{1}{2}\partial _{x}v%
\end{array}%
\right) \left(
\begin{array}{c}
2+2x \\
v+xv%
\end{array}%
\right) .  \label{3.4}
\end{equation}%
In order to integrate it, notice that
\begin{equation*}
\left(
\begin{array}{cc}
b & 0 \\
a & b%
\end{array}%
\right) \left(
\begin{array}{cc}
0 & \frac{1}{4}\partial _{x}^{3}+\frac{1}{2}u\partial _{x}+\frac{1}{2}%
\partial _{x}u \\
\frac{1}{4}\partial _{x}^{3}+\frac{1}{2}u\partial
_{x}+\frac{1}{2}\partial
_{x}u & \frac{1}{2}v\partial _{x}+\frac{1}{2}\partial _{x}v%
\end{array}%
\right) \left(
\begin{array}{c}
a \\
b%
\end{array}%
\right)
\end{equation*}%
\begin{equation*}
\ \ \ \ \ \ \ \ \ \ \ \ \ \ \ \ \ \ \ \ \ \ \ \ \ \ \ \ \ \ \ \ \
\ \ \ \ \ \ \ \ \ \ \ \ \ \ \ \ \ \ =\left(
\begin{array}{c}
\left(
\frac{1}{4}bb_{xx}-\frac{1}{8}b_{x}^{2}+\frac{1}{2}ub^{2}\right)
_{x}
\\
\left( \frac{1}{4}ab_{xx}+\frac{1}{4}ba_{xx}-\frac{1}{4}a_{x}b_{x}+uab+\frac{%
1}{2}vb^{2}\right) _{x}%
\end{array}%
\right) .
\end{equation*}%
Passing to new independent variable $y=x+1$ and substituting $a=2y$ and $%
b=yv=z$, integration of (\ref{3.4}) gives a pair of equations
\begin{align*}
0& =zz_{yy}-\frac{1}{2}z_{y}^{2}+2uz^{2}+\delta , \\
0& =\frac{1}{2}yz_{yy}-\frac{1}{2}z_{y}+2yuz+\frac{1}{2}vz^{2}+\frac{1}{2}%
\beta .
\end{align*}%
Multiplying the first equation by $y$, the second one by $-z$ and
adding them we obtain a single equation
\begin{equation*}
yzz_{yy}=yz_{y}^{2}-zz_{y}+\frac{1}{y}z^{4}+\frac{1}{2}\beta
z+\alpha y
\end{equation*}%
which, after transformation
\begin{equation*}
z=\tau q,\ \ \ \ y=\frac{1}{2}\tau ^{2},
\end{equation*}%
turns into
\begin{equation}
\tau qq_{\tau \tau }=\tau q_{\tau }^{2}-qq_{\tau }+4\tau
q^{4}+\beta q+\delta \tau ,  \label{3.5}
\end{equation}%
i.e. $P_{III}$ (\ref{1.2}) with $\alpha =0$ and $\gamma =4$.

Let us conclude this letter with an open question, whether one can
relate deformed DWW equations (\ref{2.12a})-(\ref{2.12c}) to some
non-isospectral problems, according to the idea developed in
\cite{J1}.

\label{lastpage}

\begin{thebibliography}{99}

\bibitem{A1} Bracken A.J., Bass L., Rogers C.,B\"{a}cklund flux quantization in a model of electrodiffusion based on Painlev\'{e} II, {\it J. Phys. A: Math. Theor.},
 2012, V.45, 105204 (20pp).

\bibitem{A2} Fokas A.S., Tanveer S., A Hele-Shaw problem and the second Painlev\'{e} transcendent, {\it Math. Proc. Cambridge Philos. Soc.},
 1998, V.124, 169--191.


\bibitem{A4} Aftalion C. A., Du Q., Pomeau Y., Dissipative flow and vortex shedding in the Painlev\'{e} boundary layer of a Bose-Einstein condensate, {\it Phys. Rev. Lett.},
2003, V.91, 090407 (4pp)

\bibitem{C} Clarkson P. A. and Kruskal M. D., New similarity reductions of the Boussinesq equation, {\it J. Math. Phys.}, 1989, V.30,
2201--2213.

\bibitem{J1} Gordoa P. R., Joshi N., Pickering A., On a generalized 2+ 1 dispersive water wave hierarchy, {\it Publ. RIMS. Kyoto
Univ.}, \textbf 2001, V.37, 327--347.

\bibitem{J} Joshi N., The second Painlev\'{e} hierarchy and the stationary KdV hierarchy, {\it Publ. RIMS. Kyoto Univ.}, 2004, V.40,
1039--1061.

\bibitem{F} Fokas A.S., Leo R. A., Martine L., Soliani G., The scaling reduction of the three-wave resonant system and the Painlev\'{e} VI equation, {\it Phys. Lett.
A}, 1986, V.115, 329--332.

\bibitem{G} Gordoa P.R., Mugan U., Pickering A.,  Generalized scaling reductions and Painlev\'{e} hierarchies, {\it Appl. Math. Comput.} 2013, V.%
219, 8104--8111.

\bibitem{Co} Conte R., Grundland A.M., Musette M., A reduction of the resonant three-wave interaction to the generic sixth Painlev\'{e} equation, {\it J.Phys. A: Math.
Gen.}, 2006, V.39, 12115--12127.

\bibitem{M} Murata M., J. Two-component soliton systems and the Painlev\'{e} equations, {\it Phys. A: Math. Gen.}, 2008, V.41, 365205
(17pp).

\bibitem{Ar} Aratyn H., Gomes J.F., Ruy D.V., Zimerman A.H., A symmetric system of mixed Painlev\'{e}  III--V equations and its integrable origin, {\it J. Phys. A:
Math. Theor.}, 2016, V.49, 045201 (21pp).

\bibitem{A} Antonowicz M., Fordy A., Factorisation of energy dependent Schr\"{o}dinger operators: Miura maps and modified systems, {\it Commun. Math. Phys.}, 1989, V.124,
465--486.

\end{thebibliography}
\end{document}